\documentclass[aps,preprintnumbers,superscriptaddress,showpacs]{revtex4}
\usepackage{epsfig}
\usepackage{psfrag}
\usepackage{amsfonts}
\usepackage{graphicx}
\usepackage{dcolumn}
\usepackage{bm}

\begin{document}

\title{$R^2$ corrections to the jet quenching parameter}

\author{Zi-qiang Zhang}
\email{zhangzq@cug.edu.cn} \affiliation{School of mathematical and
physics, China University of Geosciences, Wuhan 430074, China}

\author{De-fu Hou}
\email{houdf@mail.ccnu.edu.cn} \affiliation{Key Laboratory of
Quark and Lepton Physics (MOE), Central China Normal University,
Wuhan 430079,China}

\author{Yan Wu}
\email{yan.wu@cug.edu.cn} \affiliation{School of mathematical and
physics, China University of Geosciences, Wuhan 430074, China}

\author{Gang Chen}
\email{chengang1@cug.edu.cn} \affiliation{School of mathematical
and physics, China University of Geosciences, Wuhan 430074, China}

\begin{abstract}
A calculation of the $R^2$ corrections to the jet quenching
parameter from AdS/CFT correspondence is presented. These
corrections are related to curvature-squared corrections in the
corresponding gravity dual. It is shown that the corrections will
increase or decrease the jet quenching parameter depending on the
coefficients of the high curvature terms.

\end{abstract}
\pacs{12.38.Lg, 12.38.Mh, 11.25.Tq }

\maketitle
\section{Introduction}

The experiments of ultrarelativistic nucleus-nucleus collisions at
RHIC and LHC have produced a strongly-coupled quark-gluon
plasma(sQGP). One of the interesting properties of the sQGP is jet
quenching phenomenon: due to the interaction with the medium, high
energy partons traversing the medium are strongly quenched. This
phenomenon is usually characterized by the jet quenching parameter
(or transport coefficient) $\hat q$ which describes the average
transverse momentum square transferred from the traversing parton,
per unit mean free path \cite{RB,XN}.

AdS/CFT duality can explore the strongly coupled $\mathcal N=4$
supersymmetric Yang-Mills(SYM) plasma through the correspondence
between the type IIB superstring theory formulated on AdS$_5\times
S^5$ and $\mathcal N=4$ SYM in four dimensions
\cite{Maldacena:1997re,Gubser:1998bc,MadalcenaReview}. Therefore
some quantities for instance the jet quenching parameter can be
studied.

By using the AdS/CFT correspondences, H.Liu et al \cite{liu} have
calculated the jet quenching parameter for $N=4$ SYM theory
firstly. Interestingly the magnitude of $\hat q_{SYM}$ turns out
to be closer to the value extracted from RHIC data \cite{K.J,A.D}
than pQCD result for the typical value of the 't Hooft coupling,
$\lambda\simeq 6\pi$, of QCD. This proposal has attracted lots of
interest. Thus, after \cite{liu} there are many attempts to
addressing jet quenching parameter \cite{KB,ZQ,AF1,NA}. However,
in \cite{liu} the jet quenching parameter was studied only in
infinite-coupling case. Therefore, it is very interesting to
investigate the effect of finite-coupling corrections. The purpose
of the present work is to study $R^2$ corrections to the jet
quenching parameter.

The organization of this paper is as follows. In the next section,
the result of jet quenching parameter from dual gauge theory is
briefly reviewed \cite{liu}. In section 3, we consider the $R^2$
corrections to the jet quenching parameter. The last part is
devoted to conclusion and discussion.

\section{$\hat q$ from dual gauge theory}

The eikonal approximation relates the jet quenching parameter with
the expectation value of an adjoint Wilson loop $W^A[{\cal C}]$
with ${\cal C}$ a rectangular contour of size $L\times L_-$, where
the sides with length $L_-$ run along the light-cone and the limit
$L_-\to\infty$ is taken in the end. Under the dipole
approximation, which is valid for small transverse separation $L$,
the jet-quenching parameter $\hat{q}$ defined in Ref \cite{RB} is
extracted from the asymptotic expression for $TL<<1$
\begin{equation}
<W^A[{\cal C}]> \approx \exp [-\frac{1}{4\sqrt{2}}\hat{q}L_-L^2],
\label{jet}
\end{equation}
where $<W^A[{\cal C}]>\approx <W^F[{\cal C}]>^2$ with $<W^F[{\cal
C}]>$ the thermal expectation value in the fundamental
representation.

Using AdS/CFT correspondence, we can calculate $<W^F[{\cal C}]>$
according to:
\begin{equation}
<W^F[{\cal C}]>\approx\exp[-S_I] \label{WF},
\end{equation}
where $S_I$ is the regulated finite on-shell string world-sheet
action whose boundary corresponds to the null-like rectangular
loop ${\cal C}$.

By using such a strategy, they find that \cite{liu}
\begin{equation}
\hat
q_{SYM}=\frac{\pi^{3/2}\Gamma(\frac{3}{4})}{\Gamma(\frac{5}{4})}\sqrt{\lambda}T^3\approx7.53\sqrt{\lambda}T^3,
 \label{q}
\end{equation}
where $T$ is the temperature of ${\cal N}=4$ SYM plasma.

\section{$R^2$ corrections to $\hat q_{SYM}$ }

We now consider the $R^2$ corrections to $\hat q_{SYM}$.
Restricting to the gravity sector in $AdS_5$, the leading order
higher derivative corrections can be written as \cite{SM}
\begin{equation}
S=\frac{1}{16\pi G_N}\int
d^5x\sqrt{-g}[R-2\Lambda+l^2(\alpha_1R^2+\alpha_2R_{\mu\nu}R^{\mu\nu}+\alpha_3R_{\mu\nu\rho\sigma}R^{\mu\nu\rho\sigma})],\label{action}
\end{equation}
where $\alpha_i$ are arbitrary small coefficients and the negative
cosmological constant $\Lambda$ relates the AdS space radius $l$
by
\begin{equation}
\Lambda=-\frac{6}{l^2}.
\end{equation}

The black brane solution of $AdS_5$ space with curvature-squared
corrections is given by
\begin{equation}
ds^2=-f(r)dt^2+{\frac{r^2}{l^2}}d\vec{x}^2+\frac{1}{f(r)}dr^2
\label{metric}, \label{metric1}
\end{equation}
where
\begin{equation}
f(r)=\frac{r^2}{l^2}(1-\frac{r_0^4}{r^4}+\alpha+\beta\frac{r_0^8}{r^8}
) \label{f1},
\end{equation}
with
\begin{equation}
\alpha=\frac{2}{3}(10\alpha_1+2\alpha_2+\alpha_3),\qquad
\beta=2\alpha_3 \label{afa},
\end{equation}
where r denotes the radial coordinate of the black brane geometry,
$r=r_0$ is the horizon, and $l$ relates the string tension
$\frac{1}{2\pi\alpha^\prime}$ to the 't Hooft coupling constant by
$\frac{l^2}{\alpha^\prime}=\sqrt{\lambda}$.

The temperature of ${\cal N}=4$ SYM plasma with the $R^2$
corrections is given by
\begin{equation}
T_1=\frac{r_0}{\pi
l^2}(1+\frac{1}{4}\alpha-\frac{5}{4}\beta)=T(1+\frac{1}{4}\alpha-\frac{5}{4}\beta)
\label{T}.
\end{equation}

In terms of the light-cone coordinate $x^\mu=(r,x^\pm,x_2,x_3)$,
the metric (\ref{metric}) becomes
\begin{equation}
ds^2=-[\frac{r^2}{l^2}+f(r)]dx^+dx^-+\frac{r^2}{l^2}(dx_2^2+dx_3^2)+\frac{1}{2}[\frac{r^2}{l^2}-f(r)][(dx^+)^2+(dx^-)^2]+\frac{dr^2}{f(r)}.
 \label{metric1}
\end{equation}

The world sheet in the bulk can be parameterized by $\tau$ and
$\sigma$, then the Nambu-Goto action for the world-sheet becomes
\begin{equation}
S=\frac{1}{2\pi\alpha^\prime}\int d\tau
d\sigma\sqrt{detg_{\alpha\beta}}. \label{NG}
\end{equation}

We set the pair of quarks at $x_2=\pm\frac{L}{2}$ and choose
$x^-=\tau,x_2=\sigma$. In this setup, we can ignore the effect of
$x^-$ dependence of the world-sheet, implying
$x_3=const,x^+=const$, then the string action is given by
\begin{equation}
S=\frac{\sqrt{2}r_0^2L_-}{2\pi\alpha^\prime
l^2}\int_0^{\frac{L}{2}}d\sigma\sqrt{1+\frac{{r^\prime}^2l^2}{fr^2}}
\label{NG1},
\end{equation}
with $r^\prime=\partial_\sigma r$.

The equation of motion for $r(\sigma)$ reads
\begin{equation}
{r^\prime}^2=\gamma^2\frac{r^2f}{l^2} \label{gamma},
\end{equation}
where $\gamma$ is an integration constant. From Eq.(\ref{gamma}),
it is clear that the turning point occurs at $f=0$, implying
$r^\prime=0$ at the horizon $r=r_0$. Knowing that $r=r_0$ at
$\sigma=0$, then Eq.(\ref{gamma}) can be integrated as
\begin{equation}
\frac{L}{2}=\int_{r_0}^\infty\frac{ldr}{\gamma\sqrt{f}r}=\frac{l^2}{\gamma
r_0}\int_1^\infty\frac{dt}{\sqrt{t^4-1+\alpha
t^4+\frac{\beta}{t^4}}}=\frac{al^2}{\gamma r_0},
\end{equation}
where
\begin{equation}
a=\int_1^\infty\frac{dt}{\sqrt{t^4-1+\alpha
t^4+\frac{\beta}{t^4}}}. \label{a1}
\end{equation}
Then we can find the action for the heavy quark pair
\begin{equation}
S=\frac{\pi\sqrt{\lambda}L_-LT_1^2}{2\sqrt{2}}\sqrt{1+\frac{4a^2}{\pi^2T_1^2L^2}}
\label{NG2},
\end{equation}
where we have used the relation $r_0=\pi l^2T$ and
$\frac{l^2}{\alpha^\prime}=\sqrt{\lambda}$.

This action needs to be subtracted by the self-energy of the two
quarks, given by the parallel flat string world-sheets, that is
\begin{equation}
S_0=\frac{2L_-}{2\pi\alpha^\prime}\int_{r_0}^\infty
dr\sqrt{g_{--}g_{rr}}=\frac{a\sqrt{\lambda}L_-T_1}{\sqrt{2}}.
\end{equation}

Therefore, the net on-shell action is given by
\begin{equation}
S_I=S-S_0\approx\frac{\pi^2}{8\sqrt{2}a}\sqrt{\lambda}T_1^3L_-L^2,
\label{SI}
\end{equation}
where we have used $T_1L\ll1$.

Note that $\hat q$ is related to $<W^A[{\cal C}]>$ according to:
\begin{equation}
<W^A[{\cal C}]> \approx \exp [-\frac{1}{4\sqrt{2}}\hat
qL_-L^2]\approx\exp[-2S_I] \label{jet0}.
\end{equation}

Plugging (\ref{SI}) into (\ref{jet0}), we end up with jet
quenching parameter under the $R^2$ corrections
\begin{equation}
\hat
q_{R^2}=\frac{\pi^2}{a}\sqrt{\lambda}T_1^3=\frac{(1+\frac{1}{4}\alpha-\frac{5}{4}\beta)^3\pi^2}{\int_1^\infty\frac{dt}{\sqrt{t^4-1+\alpha
t^4+\frac{\beta}{t^4}}}}\sqrt{\lambda}T^3
 \label{q1}.
\end{equation}

We now discuss our result. It is clear that the jet quenching
parameter in (\ref{q}) can be derived from (\ref{q1}) if we
neglect the effect of curvature-squared corrections by plugging
$\alpha=\beta=0$ in (\ref{q1}). Furthermore, we compare the jet
quenching parameter under the $R^2$ corrections with its
counterpart in the case of the dual gauge theory as following
\begin{equation}
\frac{\hat q_{R^2}}{\hat
q_{SYM}}=\frac{\Gamma(\frac{5}{4})\sqrt{\pi}}{\Gamma(\frac{3}{4})}\frac{(1+\frac{1}{4}\alpha-\frac{5}{4}\beta)^3}{\int_1^\infty\frac{dt}{\sqrt{t^4-1+\alpha
t^4+\frac{\beta}{t^4}}}}.
 \label{q11}
\end{equation}

Notice that the result of Eq.(\ref{q11}) is related to $\alpha$
and $\beta$, or in other words, it is dependent of $\alpha_i$.
However, the exact interval of $\alpha_i$ has not been determined,
we only know that $\alpha_i\sim \frac{\alpha\prime}{l^2}\ll1$
\cite{MB}. In order to choose the values of $\alpha$ and $\beta$
properly, we employ the Gauss-Bonnet gravity \cite{RG} which has
nice properties that are absent for theories with more general
ratios of the $\alpha_i's$. The Gauss-Bonnet gravity is defined by
the following action \cite{BZ}
\begin{equation}
S=\frac{1}{16\pi G_N}\int
d^5x\sqrt{-g}[R-2\Lambda+\frac{\lambda_{GB}}{2}l^2(R^2-4R_{\mu\nu}R^{\mu\nu}+R_{\mu\nu\rho\sigma}R^{\mu\nu\rho\sigma})].\label{action1}
\end{equation}

Comparing (\ref{action}) with (\ref{action1}), we find
\begin{equation}
\alpha_1=\frac{\lambda_{GB}}{2},\qquad
\alpha_2=-2\lambda_{GB},\qquad
\alpha_3=\frac{\lambda_{GB}}{2}\label{afai}.
\end{equation}

Plugging (\ref{afai}) into (\ref{afa}), we have
\begin{equation}
\alpha=\lambda_{GB},\qquad\beta=\lambda_{GB},
\end{equation}
where $\lambda_{GB}$ can be constrained by causality and
positive-definiteness of the boundary energy density \cite{AF1}
\begin{equation}
-\frac{7}{36}<\lambda_{GB}\leq\frac{9}{100}.
\end{equation}

Moreover, there is an obstacle to the calculation of (\ref{q11}):
we must require that the square root in the denominator is
everywhere positive and the temperature in Eq.(\ref{T}) is also
positive. Under the above conditions, we find
\begin{equation}
-\frac{7}{36}<\alpha\leq\frac{9}{100},\qquad
0\leq\beta\leq\frac{9}{100},\qquad\alpha+\beta\geq0,
\end{equation}
then we can discuss the results of (\ref{q11}) for different
values of $\alpha$ and $\beta$ in such a range.

As a concrete example, we plot $\hat q_{R^2}/\hat q_{SYM}$ versus
$\alpha$ with different $\beta$ and $\hat q_{R^2}/\hat q_{SYM}$
versus $\beta$ with different $\alpha$ in FIG.1. In the left plot
from top to down, $\beta=0.001,0.003,0.006$, $\hat q_{R^2}/\hat
q_{SYM}$ increases at each $\beta$ as $\alpha$ increases. While in
the right plot from top to down $\alpha=0.005,0.001,-0.001$, at
each $\alpha$, $\hat q_{R^2}/\hat q_{SYM}$ is a decreasing
function of $\beta$.

\begin{figure}
\centering
\includegraphics[width=8cm]{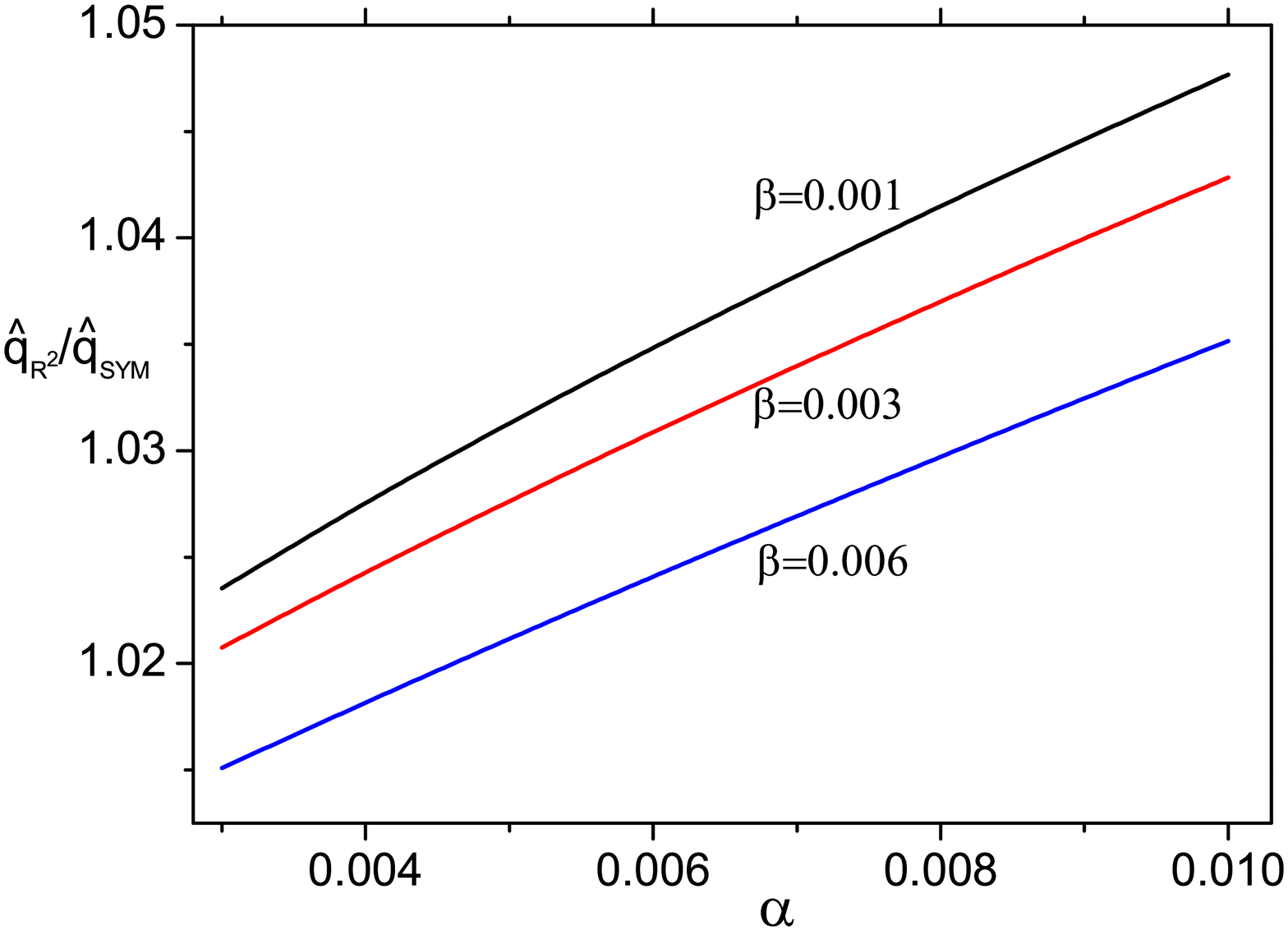}
\includegraphics[width=8cm]{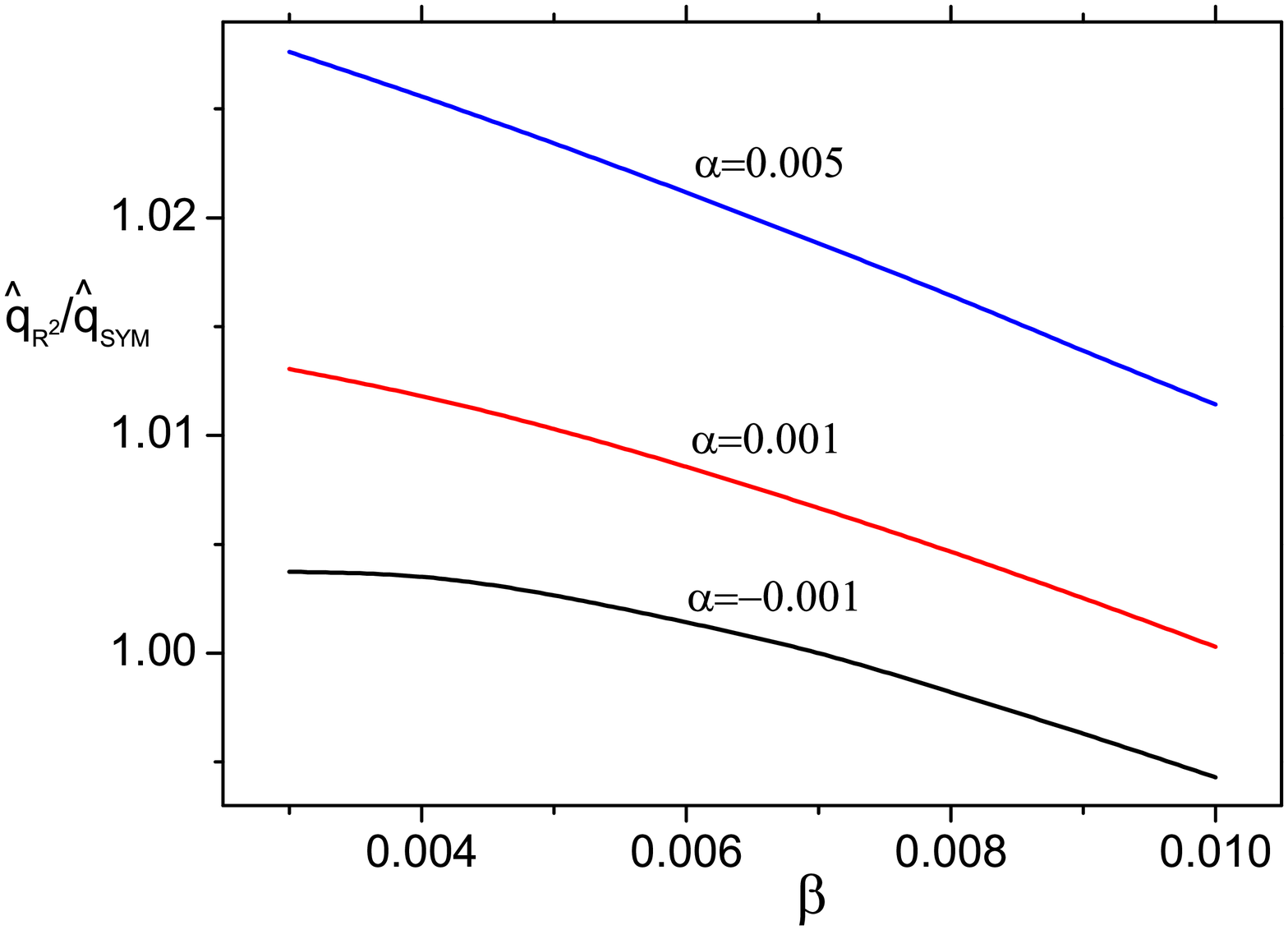} \caption{Left: $\hat
q_{R^2}/\hat q_{SYM}$ versus $\alpha$. From top to down
$\beta=0.001,0.003,0.006$. Right: $\hat q_{R^2}/\hat q_{SYM}$
versus $\beta$. From top to down $\alpha=0.005,0.001,-0.001$. }
\end{figure}

From the figure, it is clear that $R^2$ corrections can affect the
jet quenching parameter. With some chosen values of $\alpha$ and
$\beta$ in this paper, the jet quenching parameter can be larger
than or smaller than its counterpart in infinite-coupling case.

\section{conclusion and discussion}

In this paper, we have investigated the $R^2$ corrections to the
jet quenching parameter. These corrections are related to
curvature-squared corrections in the corresponding gravity dual.
It is shown that the corrections will increase or decrease the jet
quenching parameter depending on the coefficients of the high
curvature terms. Interestingly, under $R^2$ corrections, the drag
force is also larger than or smaller than that in the
infinite-coupling case \cite{KB1}. However, we should admit that
we cannot predict a result for ${\cal N}=4$ SYM because the first
higher derivative correction in weakly curved type II B
backgrounds enters at order $R^4$, and not $R^2$.

Actually, there are some other important finite-coupling
corrections to the jet quenching parameter. For example, next to
leading order corrections to (\ref{q}) due to world sheet
fluctuations suggests \cite{ZQ}
\begin{equation}
\hat q'_{SYM}=\hat q_{SYM}(1-1.97\lambda^{-1/2}).
\end{equation}

Other corrections of $O(1/N_c )$ and higher orders in
$\frac{1}{\sqrt{\lambda}}$ are also discussed in \cite{AF1,NA}.
However, $1/\lambda$ corrections on jet quenching parameter are as
yet undetermined, we hope to report our progress in this regard in
future.

\section{Acknowledgments}

We would like to thank Prof. Hai-cang Ren for useful discussions.
This research is partly supported by the Ministry of 502 Science
and Technology of China (MSTC) under the ¡°973¡± 503 Project No.
2015CB856904(4). Zi-qiang Zhang is supported by the NSFC under
Grant Nos.11547204. Gang Chen is supported by the NSFC under Grant
Nos 11475149. De-fu Hou is partly supported by NSFC under Grant
Nos. 11375070 and 11221504.



\begin{thebibliography}{0}

\bibitem{RB}
R. Baier, Y. L. Dokshitzer, A. H. Mueller, S. Peigne and D.
Schiff, {\sl "Radiative energy loss and p(T)- broadening of high
energy partons in nuclei"}, Nucl. Phys. B {\bf 484}, 265 (1997)
[hep-ph/9608322].

\bibitem{XN}
M. Gyulassy, I. Vitev, X.-N. Wang, B.-W. Zhang, {\sl "Jet
Quenching and Radiative Energy Loss in Dense Nuclear Matter"},
[nucl-th/0302077].

\bibitem{Maldacena:1997re}
J.~M.~Maldacena, {\sl "The large $N$ limit of superconformal field
theories and supergravity"}, Adv.\ Theor.\ Math.\ Phys.\  {\bf 2},
231 (1998) [Int.\ J.\ Theor.\ Phys.\  {\bf 38}, 1113 (1999)]
[hep-th/9711200].

\bibitem{Gubser:1998bc}
S.~S.~Gubser, I.~R.~Klebanov and A.~M.~Polyakov, {\sl "Gauge
theory correlators from non-critical string theory"}, Phys.\
Lett.\ B {\bf 428}, 105 (1998) [hep-th/9802109].

\bibitem{MadalcenaReview}
O. Aharony, S. S. Gubser, J. Maldacena, H. Ooguri and Y. Oz, {\sl
"Large N field theories, string theory and gravity"} Phys. Rept.
{\bf 323}, 183 (2000).

\bibitem{liu}
H. Liu, K. Rajagopal and U. A. Wiedemann, {\sl "Calculating the
jet quenching parameter from AdS/CFT"}, Phys. Rev. Lett. {\bf 97},
182301 (2006) [hep-ph/0605178].

\bibitem{K.J}
K.J. Eskola, H. Honkanen, C.A. Salgado, U.A. Wiedemann, {\sl "The
fragility of high-pT hadron spectra as a hard probe"} , Nucl.
Phys. A {\bf 747}, 511 (2005) [hep-ph/0406319].

\bibitem{A.D}
A. Dainese, C. Loizides and G. Paic, {\sl "Leading-particle
suppression in high energy nucleus-nucleus collision"}, Eur. Phys.
J. C {\bf 38} 461 (2005). [hep-ph/0406201].

\bibitem{KB}
K. B. Fadafan,  {\sl "Charge effect and finite 't Hooft coupling
correction on drag force and Jet Quenching Parameter"}, Eur. Phys.
J. C {\bf 68} 505 (2010). [hep-th/0809.1336].

\bibitem{ZQ}
Z. -Q. Zhang, D. -F. Hou and H. -C. Ren, {\sl "The finite 't Hooft
coupling correction on the jet quenching parameter in a N=4 Super
Yang-Mills Plasma"}, JHEP {\bf 1301} 032 (2013).
[hep-th/1210.5187].

\bibitem{AF1}
A. Ficnar, S. S. Gubser and M. Gyulassy, {\sl "Shooting String
Holography of Jet Quenching at RHIC and LHC"}, Phys.Lett.
B{\bf738} 464(2014). [hep-ph/1311.6160].

\bibitem{NA}
N. Armesto, J. D. Edelstein and J. Mas, {\sl "Jet quenching at
finite `t Hooft coupling and chemical potential from AdS/CFT"},
JHEP {\bf 0609} 039 (2006). [hep-ph/0606245].

\bibitem{SM}
S. M. Carroll, {\sl "Spacetime and geometry: An introduction to
general relativity"}, San Francisco, USA: Addison-Wesley (2004)
513 p

\bibitem{MB}
M. Brigante, H. Liu, R. C. Myers, S. Shenker, S.Yaida {\sl
"Viscosity Bound Violation in Higher Derivative Gravity"},
Phys.Rev.D {\bf 77} 126006(2008). [hep-th/0712.0805].

\bibitem{RG}
R. G. Cai, {\sl "Gauss-Bonnet black holes in AdS spaces"}, Phys.
Rev. D 65, 084014 (2002) [hep-th/0109133].

\bibitem{BZ}
B. Zwiebach, {\sl "Curvature squared terms and string theories"},
Phys. Lett. B {\bf 156}, 315 (1985).

\bibitem{KB1}
K. B. Fadafan,  {\sl "$R^2$ curvature-squared corrections on drag
force"}, JHEP {\bf 0812} 051 (2008). [hep-th/0803.2777].



\end{thebibliography}
\end{document}